\documentclass[preprint2]{aastex}

\begin{document}
\title{Efficiency Measurements and Installation of a New Grating for the
OSIRIS Spectrograph at Keck Observatory}
\author{Etsuko Mieda\altaffilmark{1, 2, a}, Shelley
  A. Wright\altaffilmark{1, 2}, James E. Larkin\altaffilmark{3}, James
  R. Graham\altaffilmark{2, 5}, Sean M. Adkins\altaffilmark{4}, James
  E. Lyke\altaffilmark{4}, Randy D. Campbell\altaffilmark{4},
  J{\'e}r{\^o}me Maire\altaffilmark{2}, Tuan Do\altaffilmark{2}, and
  Jacob Gordon\altaffilmark{1, 2}}
\affil{\altaffilmark{1}Department of Astronomy \& Astrophysics, University of Toronto,
  ON, CANADA, M5S 3H4} 
\affil{\altaffilmark{2}Dunlap Institute for Astronomy \&
  Astrophysics, University of Toronto, ON, CANADA, M5S 3H4}
\affil{\altaffilmark{3}Division of Astronomy and Astrophysics, University of California,
  Los Angeles, CA, USA, 90095}
\affil{\altaffilmark{4}W. M. Keck Observatory, HI,USA, 96743}
\affil{\altaffilmark{5}Astronomy Department, University of California,
  Berkeley, CA, USA, 94720}
\altaffiltext{a}{mieda@astro.utoronto.ca}


\begin{abstract}
OSIRIS is a near-infrared integral field spectrograph operating behind
the adaptive optics system at W. M. Keck Observatory. While OSIRIS has
been a scientifically productive instrument to date, its
sensitivity has been limited by a grating efficiency that is less than
half of what was expected. The spatially averaged efficiency of the old
grating, weighted by error,
is measured to be 39.5 $\pm$ 0.8 \% at $\lambda$ = 1.310 $\mu$m,
with large field dependent variation of 11.7 \% due to efficiency
variation across the grating surface. Working with a new vendor, we developed
a more efficient and uniform grating with a weighted average efficiency at 
$\lambda$ = 1.310 $\mu$m of 78.0 $\pm$ 1.6 \%, with field variation of only 2.2 \%. 
This is close to double the average efficiency and five times less variation across the field.
The new grating was installed in December 2012, and on-sky
OSIRIS throughput shows an average factor of 1.83 improvement in sensitivity
between 1 and 2.4 microns.  
We present the development history, testing, and implementation of this
new near-infrared grating for OSIRIS and report the comparison with the
predecessors. The higher sensitivities are already having a large impact
on scientific studies with OSIRIS.
\end{abstract}

\section{Introduction}
In the last decade, the combination of a near-infrared integral field
spectrograph (IFS) and 
adaptive optics (AO) has proven to be crucial in a range of astronomical
studies from our solar system to galaxies in the early
universe. Some example observations include the sulphur dioxide distribution on
one of the Galilean moons, Io \citep{laver:09}, morphology of novae
ejecta \citep{lyke:09}, the
atmosphere of an extrasolar gas giant planets \citep[e.g.][]{barman:11, konopacky:13}, the
crowded stellar fields of the Galactic Center \citep[e.g.][]{trippe:08,
  do:09:1, do:13}, AGN \citep[e.g.][]{davies:07, mcconnell:11,
  contini:12}, and high redshift galaxies 
\citep[e.g.][]{forster:06, law:09, wright:09, wisnioski:11}. IFSs are
also aimed to be the first eight instruments for the
next generation of extremely large telescopes, such as IRIS on TMT
\citep{larkin:10}, HARMONI on E-ELT \citep{thatte:10}, and GMTIFS on GMT
\citep{mcgregor:12}. 

OSIRIS (OH-Suppressing Infrared Imaging Spectrograph)  \citep{larkin:03,
  larkin:06}, a moderate 
spectral resolution (R $\sim$ 3800) diffraction limited IFS for the AO
system at W. M. Keck Observatory, is one of a handful of IFS instruments
in use with AO systems worldwide today. It was the first diffraction
limited IFS instrument to use a lenslet array as the sampling element on the sky and
has plate scales ranging from 0.02" to 0.1" per spaxel\footnote{Spectrum
  of each spatial element.}. OSIRIS' optics and lenslet
array produce low non-common path error (\textless 30 nm rms), a factor
of approximately three times 
less than any other IFS, preserving the diffraction limited
point spread function of the Keck AO system
\citep{wizinowich:06}.

OSIRIS was designed with a single fixed diffraction grating to ensure
spectral stability and make data reduction possible with very dense spectral
packing on the detector (only two pixel spacing between spectra). The grating 
is used in multiple orders ($m$) to cover
traditional near-infrared wavebands: $K$ ($\lambda_{cen}$ = 2.2 $\mu$m) is
sampled in $m = -3$, $H$ ($\lambda_{cen}$ = 1.6 $\mu$m) in $m = -4$, $J$
($\lambda_{cen}$ = 1.3 $\mu$m) in $m = -5$, and $Z$ ($\lambda_{cen}$ = 1.1
$\mu$m) in $m = -6$.

While OSIRIS has been a productive instrument to date, its performance
has been limited by sensitivity, which is approximately 50 \% lower than
its design prediction, particularly at shorter wavelengths ($Z$ and $J$ bands). Through
our team's investigation, this performance limitation has been
determined to be due to the quality of the spectrograph's diffraction
grating. Since 2009 our team actively pursued acquiring a new grating
for the OSIRIS spectrograph. In 2011, we began to work with the Bach Research
Corporation, Boulder, CO, to fabricate a new, more efficient grating for OSIRIS. 
Our goal was to improve the grating performance sufficiently to double the
signal to noise ratio for detector limited observations. 

In this paper, we describe our acquisition and testing of a new
grating.
In \textsection \ref{history}, we summarize the history of the OSIRIS
spectrograph grating. In \textsection
\ref{measurement} we describe the laboratory setup used to measure the
grating efficiency and the results of those measurements. In \textsection
\ref{installation}, we report on the December 2012 installation of the
new grating. In \textsection \ref{onsky}, we discuss the on-sky
performance of OSIRIS with the new grating.
For interested in equipment characterization, Appendix
\ref{sec_camera} and \ref{sec_laser} describe camera and laser diode
characterization processes in detail. For the user of OSIRIS or other
IFS instruments, we introduce the OSIRIS data reduction pipeline and the
modifications made after installation of the new grating in Appendix
\ref{pipeline}.
 
\section{History of OSIRIS Grating}
\label{history}
The OSIRIS spectrograph grating is a unique and unusual single fixed
diffraction grating that has a coarse ruling of 27.93 grooves per
mm at a shallow blaze angle of 5.76$^\circ$. The specifications of the
grating are listed in Table \ref{spec}. 
\begin{deluxetable}{ll}
\tablecolumns{2}
\tablewidth{0pc}
\tablecaption{OSIRIS Grating Specification}
\setlength{\tabcolsep}{0.07in}
\tablehead{
\colhead{Parameter} & \colhead{Value}
}
\startdata
Size & 275 $\times$ 220 $\times$ 50 mm \\
Line Spacing & 27.93 lines/mm \\
Blaze Angle & 5.76$^\circ$ \\
Clear Aperture & 205 $\times$ 230 mm (min.) \\
Surface Irregularity & 150 nm RMS \\
Surface & Gold coating on aluminum substrate \\
\enddata \label{spec}
\end{deluxetable}

The grating design was done by Richardson Gratings, Rochester, NY, in collaboration
with SSG Precision Optronics, Inc., Wilmington, MA, the designers
and fabricators of the OSIRIS collimator and camera three-mirror
anastigmats.

Over the time OSIRIS has been in service at Keck Observatory, we have
installed three different gratings. They are summarized in Table
\ref{g_summary}.
\begin{deluxetable}{cccc}
\tablecolumns{4}
\tablewidth{0pc}
\tablecaption{Summary of All Three OSIRIS Gratings}
\setlength{\tabcolsep}{0.07in}
\tablehead{
\colhead{Grating} & \colhead{Manufacturer} & \colhead{Service Duration}
& \colhead{$J$ Efficiency}
}
\startdata
G1 & Diffraction Products, Inc. & Feb 2005 - May 2005 & $\sim$ 15 \% \\
G2 & Diffraction Products, Inc. & Jun 2005 - Dec 2012 & 39.5 \% \\
G3 & Bach Research Corporation & Jan 2013 - present & 78.0 \% \\
\enddata \label{g_summary}
\end{deluxetable}

Originally, SSG manufactured two large aluminum grating blanks and provided these
to Richardson Gratings for ruling. However, this first option for ruling
was abandoned by them due to the large amount of tool pressure that would be required. 
A new vendor, Diffraction Products,
Inc., Woodstock, IL, then agreed to take on the challenge of ruling this
very coarse grating directly into a pure gold coating placed
on the SSG aluminum blank. The resulting grating is identified as G1 in
Table \ref{g_summary}. 

During laboratory testing of OSIRIS in October 2004, it was determined
that G1 had a slightly varying, incorrect (6.2$^\circ$ instead of
5.76$^\circ$) blaze angle. At high order, this puts the majority
of the light into the wrong order and off the field of the
detector. Efficiencies in the $Z$ and $J$ band were below 20\% and
even in the K-band was below 30\%. Due to time constraints, OSIRIS was
shipped to the telescope with this imperfect grating while a replacement
was ordered. Diffraction products significantly improved
their process and a replacement grating with the correct blaze angle
(called G2) was installed in OSIRIS in June 2005.

\begin{figure}[!h]
\plotone{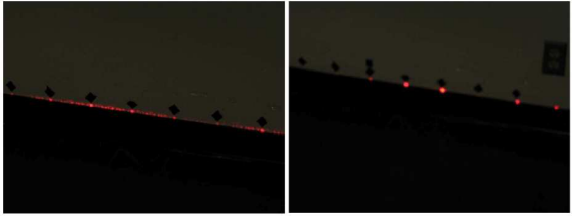}
\caption{A photograph of diffraction spots at the Keck
  Observatory in June 2005. It shows a HeNe laser at 632.8 nm being
  diffracted by G1 (left) and G2 (right). The locations of diffraction
  spots were marked with electrical tape. Note that G1 produces
  dramatic light loss between the orders compared to G2.}
\label{first_second_grating}
\end{figure}
Figure \ref{first_second_grating} is a photograph of diffraction spots
on a wall by G1 (left) and G2 (right) at the Keck Observatory in June
2005. The left image shows scattered light between the different orders due
to incomplete ruling. This shows the
improvement of the grating quality visually. The throughput measurement
at the time of the servicing mission showed a gain of a factor of three
to four in $J$ band with a smaller gain at longer wavelengths.

\begin{figure}[!h]
\plotone{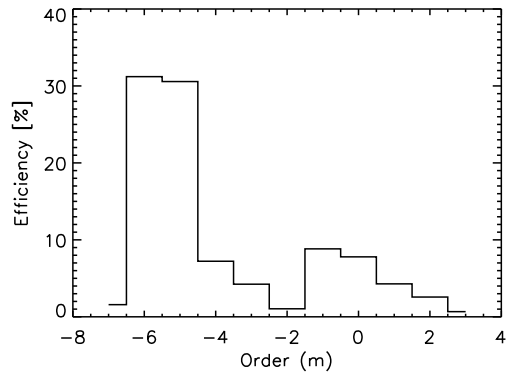}
\caption{Measurements of absolute efficiency by order for G2 at the Keck
  summit in 2011. The blaze wavelength
  is 6.5 $\mu$m, so the 5$^{th}$ order should have the maximum power at $\sim$60 to
  70 \%.}
\label{second_efficiency}
\end{figure}

\begin{figure}[!h]
\plotone{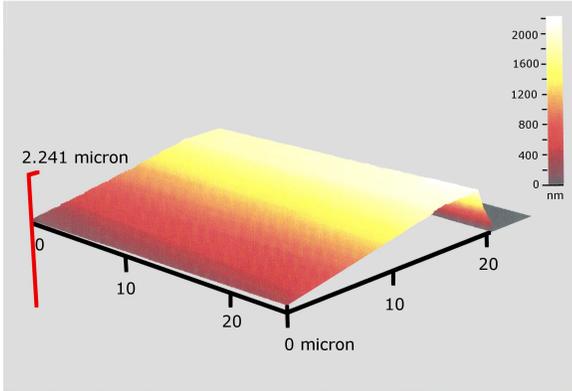}
\caption{AFM scan of one of the facets of G2 made by Diffraction
  Products, illustrating the curved profile on
  the facet, which decreases the overall sensitivity in each order.}
\label{afm_scan}
\end{figure}

Unfortunately, even with G2, the throughput was still $\sim$ 50 \% of
what was expected. This was later confirmed by our team during an October 2009 
servicing mission. G2 was removed from OSIRIS, and its efficiency was
measured at Keck Observatory using a
1.310 $\mu$m laser (close to 5th order expected blaze wavelength) and an
infrared camera. The resulting absolute efficiency
measurements are shown in Figure \ref{second_efficiency}. 
This was also verified with atomic force microscope (AFM) scans of
G2. An AFM scan of one of the grating facets is shown in
Figure \ref{afm_scan}. The grating facet shows a flat spot at the edge
of the ruling, and the profile on the primary facet has at
least two distinct angles.
The effect of the curved profile on the steep side of the profile is to
distribute some of the energy from the
expected order into adjacent orders. The expected effect of the flat
spot in the facets is that it causes some of the light to be scattered
across all of the orders. We do not know if the same facet profile occurs
throughout the grating, but both of these
effects are clearly visible in our efficiency plot in Figure
\ref{second_efficiency}. Most likely, G2 has generally poor quality
groove shapes like Figure \ref{afm_scan}. If all of the
energy between orders -5 and -6 in Figure \ref{second_efficiency} were
concentrated in the expected order ($m$ = -5), the
efficiency at 1.310 $\mu$m would be $\textgreater$ 60 \% as expected from the
specifications. 

At this point, the OSIRIS team began a search for a new vendor to manufacture a
better quality grating. Bach Research Corporation, formed
by the founding members of the Hyperfine company, began making
custom astronomical gratings, and we selected them in 2011 to
begin the process of ruling a new grating on the original SSG blank.

The first and second OSIRIS gratings were directly ruled into a gold
coating applied to a machined one piece aluminum grating substrate and
grating mount. Rather than directly ruling into the grating substrate, Bach
Research suggested
that we replicate the grating onto the aluminum substrate and then coat
the replica
with gold. The one piece machined aluminum grating mount and substrate
is an expensive component to machine, so we made use of the spare
grating mount used during the first attempt of the grating by
Diffraction Products. To produce the new grating, Bach Research removed 
the coating from G1 mount and re-polished it. This provided
a new surface to apply a new ruling using a replication process. The
fabrication was performed in two steps by Bach Research Corporation:
\begin{itemize}
\item[(1)] A new master grating was ruled onto a Zerodur substrate (a
  glass-ceramic composite material produced by Schott AG).
\item[(2)] The new master was used for replication of the grating on a
  new coating on the spare substrate (called G3).
\end{itemize}

\begin{figure}[!h]
\plotone{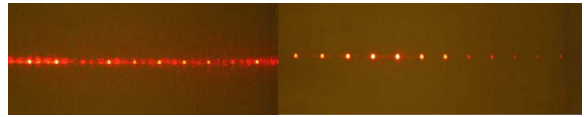}
\caption{Diffraction of HeNe laser at 632.8 nm using G1 (left) and the test ruling
  made by Bach Research (right) compared at Back Research.}
\label{first_third_grating}
\end{figure}

As part of the contract discussions for the new grating, Bach Research
made a demonstration test ruling on a small 5 mm $\times$ 100 mm long
substrate. A comparison of a diffraction of a HeNe laser at 632.8 nm
with both this substrate and
G1, taken at Bach Research is shown in Figure
\ref{first_third_grating}. 
The left image is the diffraction spots produced by G1 and the right
image is by the test ruling. The light diffracted by
the test ruling is well concentrated in spots while G1
smears the light in the direction of dispersion.

One of the important challenges encountered in the manufacture of the
previous gratings was that the grating maker could only evaluate the
grating performance using a simple set up involving a HeNe laser with
visual evaluation of the resulting dispersion and relative intensities
in each order. The method used did not predict the grating's eventual
performance at infrared wavelengths. Before
installation of G3 we acquired an infrared laser
source (1.310 $\mu$m) and infrared camera and created a set-up that allowed
measurement of the grating efficiency in a reliable fashion. This
allowed us to evaluate the test ruling as well as the final grating before
it was installed in OSIRIS.

\section{Grating Efficiency Measurement}
\label{measurement}
To investigate the grating performance in a more robust manner, we measure
the direct efficiency of the grating at $\lambda$ = 1.310 $\mu$m, which
corresponds to a wavelength in the $J$ band. In this section, we
describe the measurement equipment,
measurement setup, procedure, and discuss the measurement results. 

\subsection{Measurement Equipment and Stability}
\label{setup}
For the grating efficiency measurements in infrared, we used an InGaAs
camera (Raptor Photonics OWL SW 1.7 CL-320) and a 1.310 $\mu$m laser diode
coupled to a SMF-28 fiber (a single mode fiber with a core diameter of
8.2 $\mu$m operating at 1.310 $\mu$m to 1.625 $\mu$m). The linearity of
the camera and the stability of
the laser are discussed in Appendices
\ref{sec_camera} and \ref{sec_laser}. Some preliminary tests were done on a 5
$\times$ 100 mm test ruling that Bach Research fabricated in early
2012 before G3 manufacture. In
summary, we find that 10 data number (DN) noise level in the camera, a 0.3 \%
fluctuation due to a 10$^\circ$C camera temperature change, and a 3.5 \% fluctuation in
laser diode intensity over a one hour period.

\subsection{Measurement Setup and Procedure}
\label{procedure}
\begin{figure}[!h]
\plotone{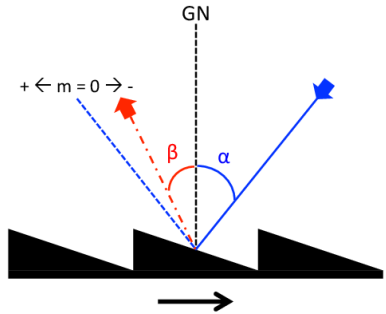}
\caption{Orientation of OSIRIS grating blaze direction (black
  arrow), incident angle $\alpha$ (blue and negative), and outgoing
  angle $\beta$
  (red, dot-dash line and positive). Negative orders are defined to
  be in the direction towards the grating normal (GN) from $m = 0$ (pure
  reflection).} 
\label{angle_config}
\end{figure}

\begin{figure}[!h]
\plotone{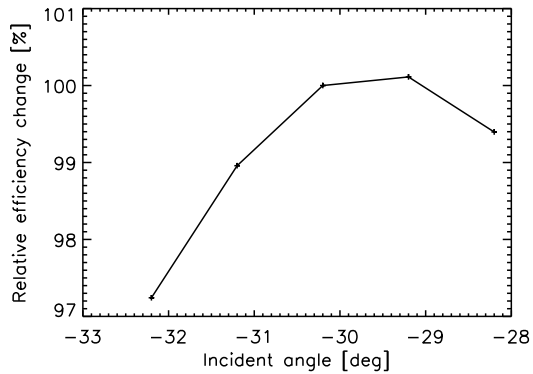}
\caption{RCWA analysis prediction of the theoretical $m = -5$ efficiency
  change at $\lambda$ = 1.310 $\mu$m due
  to the incident angle change. With a 1$^\circ$ change in the incident
  angle, the change in efficiency is less than 1\%.}
\label{eff_deg}
\end{figure}

To measure the efficiency of the grating at the same configuration as in
OSIRIS, we set
up an optical path on an optical bench, where the angle of incidence
($\alpha$) is $\alpha = -30.2^\circ$, and the angle of diffraction
($\beta$) for $m = -5$ at 1.310 $\mu$m is $ \beta = 18.4^\circ$. We
define the sign conventions and orientation for the OSIRIS grating used
throughout our measurements in Figure \ref{angle_config}.
In our setup, the incident angle is accurate
within 1$^\circ$, which is theoretically a \textless 1 \% change in $m = -5$
efficiency at 1.310 $\mu$m (Figure \ref{eff_deg}) using our Rigorous
Coupled-Wave Analysis (RCWA) \citep{moharam:81} code.
RCWA is a semi-analytic computational
method used to solve Maxwell's equations. Our code uses this method
to calculate the portion of light being diffracted into different orders
by a diffraction grating for the given grating specification,
Table \ref{spec}, and the incident angle. 

\begin{figure*}[!h]
\begin{center}
\includegraphics[width=1\textwidth]{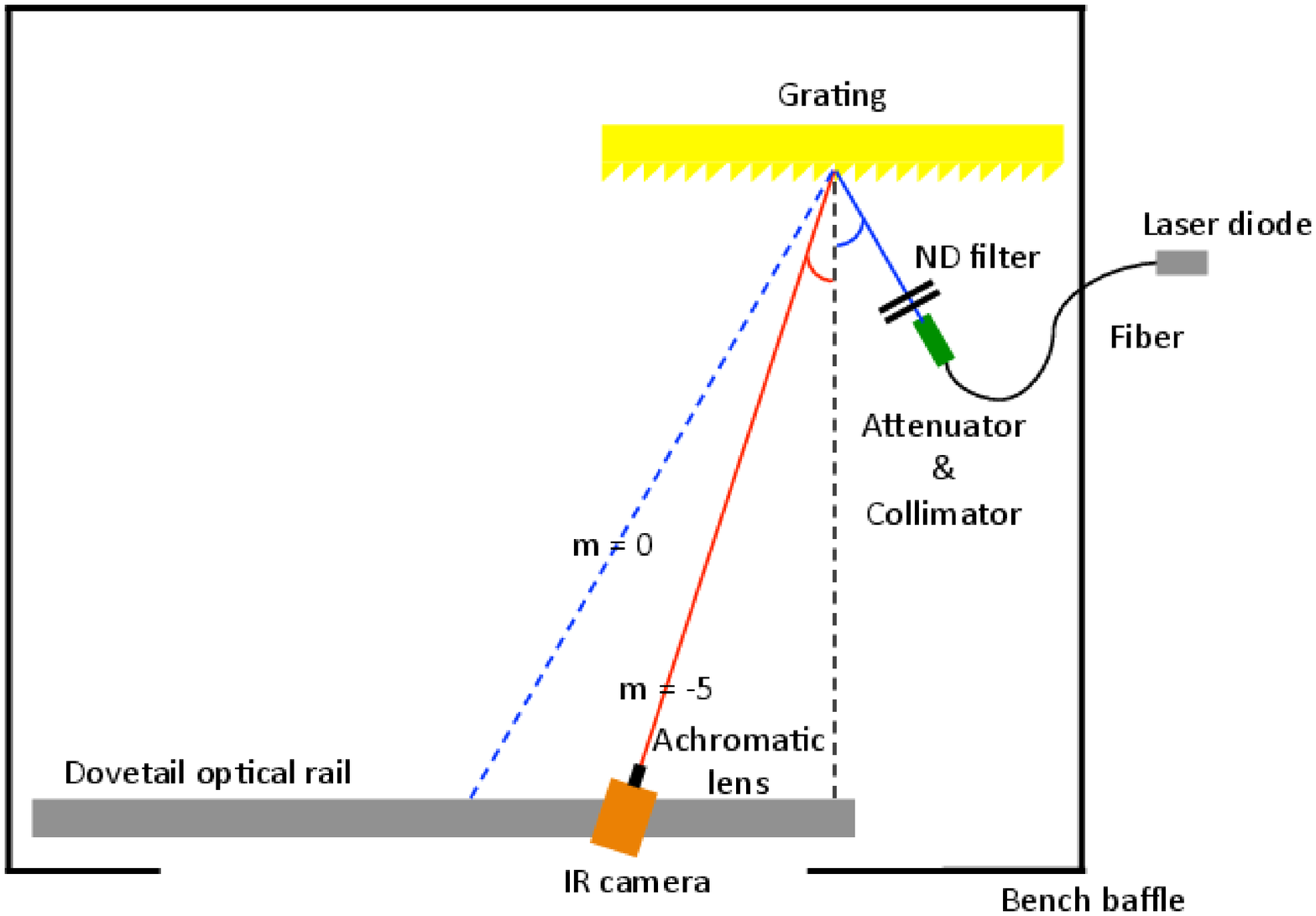}
\caption{Schematic of the efficiency measurement configuration in
  the lab (not to scale). The entire setting is covered up by an
  aluminum baffle box to control background and scattered light.}
\label{opt_config}
\end{center}
\end{figure*}

\begin{figure}[!h]
\plotone{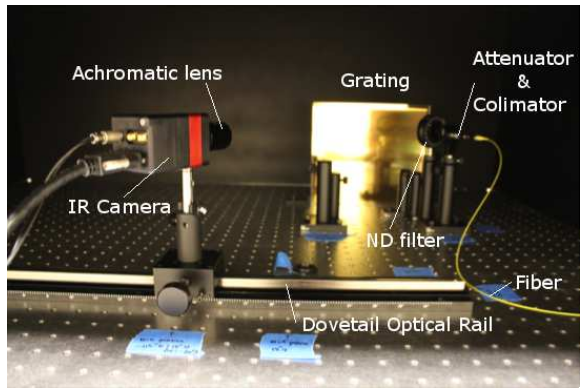}
\caption{A photo of the efficiency measurement configuration in the
  lab.}
\label{config_pic}
\end{figure}

A 1.310 $\mu$m laser diode that is coupled to a SMF-28 fiber is
connected to an attenuator and a collimator. The collimated laser beam goes
through two neutral density (ND) filters and hits the grating surface,
where it is diffracted into constituent orders. An achromatic
lens pair focuses the beam on the InGaAs
camera, which sits on a dovetail optical rail system. 
The beam's full-width-half-maximum at $m$ = -5 is about 1.7 mm on the camera. 
The schematic of
the configuration is shown in Figure \ref{opt_config},
and a photo of the setup is on Figure \ref{config_pic}.
An aluminum baffle box resides
over the entire experiment to eliminate scattered background light.
All the components, except the grating, were kept fixed onto the optical
bench until all efficiency measurements were completed. Every time we
left the lab, the grating was carefully packed and put away in a secured
location. 

The collimator focal length and the
optical path length were determined by considering the divergence angle
of the laser with the goal of keeping the final spot size well inside
the detector field of view (FOV). The
attenuator and combination of two ND filters are employed to ensure that
the final
spot on the detector is not saturated at a reasonable exposure time for
the brightest order with good signal to noise ratio on the faintest
orders. The achromatic camera lens pair is chosen so that only one spot falls
on the camera's detector at a time.

The efficiency is defined as the flux of monochromatic light diffracted
into the order being measured relative to the total flux. We measure a pure
reflection of the same light source from an un-ruled area on the
grating/test ruling (called
pure reflection) as the total efficiency. Since the un-ruled part of the
grating/test ruling is
outside of the clear aperture, and the quality of its surface is not
guaranteed, we also use the total sum of all orders (called order sum)
as a measure of the total flux as well. Efficiency measurements using
both values for total flux are presented in this paper.

A typical efficiency measurement procedure is as follows: 1) set up the
grating for the pure reflection; 2) close the bench baffle; 3) find the best
exposure time and measure the pure
reflection with the laser on; 4)  measure the pure reflection with the laser
off; 5) open the baffle; 6) set up the grating to place the first order
to be measured in the camera's FOV;
7) close the baffle; 8) find the best exposure time and measure the flux
with laser on and off; 9) move the IR camera to the next order;
and repeat step 8 and 9 until all orders are measured. In these tests,
we measured $m = -13$ to $m = 8$. It takes about
one hour to complete this procedure. Since we know that the polarization
state of the fiber does not change in one hour (Appendix
\ref{sec_laser}) but moving the fiber changes the polarization state of
the beam, we were very careful not to touch the fiber during the
entire procedure.

\begin{figure}[!h]
\plotone{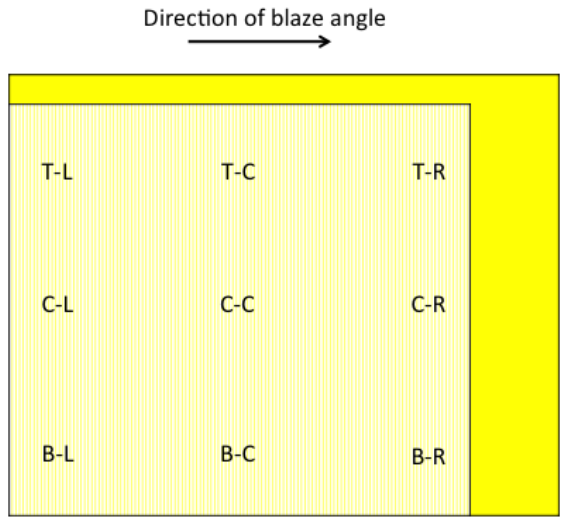}
\caption{Diagram showing the location of the grating efficiency
  measurements. The light yellow area is the region of the ruled area,
  205 $\times$ 230 mm. The bright yellow area is the overall area of the
  substrate. The arrow indicates the direction of blaze angle.}
\label{eff_loc}
\end{figure}

The ruled area of the OSIRIS grating is 205 $\times$ 230 mm. To assess
the spatial dependence of the efficiency, measurements are made at nine
locations (3 by 3 configuration) across the grating surface as
illustrated in Figure \ref{eff_loc}. 

After the efficiency at one location is measured, we move the grating
sideways or change the height of the stage, where the grating sits, to
move to the next location. The grating surface is kept
parallel to the dovetail optical rail, and the optical path length
is kept the same for all measurement. This allows us to keep the setup fixed as much
as possible. 

For each grating order, we acquire 100 frames with an additional 100
background frames. The 100 background frames are median combined to make
a master background and subtracted from each science frame. Then 100
background subtracted frames are median combined and divided by the
exposure time to make the final reduced image.
To conserve the optical alignment, we do not take flat frames. 
A histogram of a normalized flat field taken during testing shows a
normal distribution with a standard deviation of
0.024. Instead of applying flat fielding to the final reduced image, we
include the flat field fluctuation of 2.4 \% as a part of measurement
uncertainties. 
 
A two dimensional
Gaussian function is fitted to the final reduced image to find the spot
center, and the total flux in the spot is obtained by summing up the
counts in the biggest circle that can be fit in the final reduced image
centered at the spot center. A more detailed discussion of this method
is found in Appendix \ref{justification}.

There were two efficiency requirements defined by Keck Observatory and
our team which the new grating (G3) had to meet in
order to be eligible for installation in OSIRIS:
\begin{itemize}
\item[(A)] Global efficiency requirement: On average, the new
grating has to be at least 50 \% more efficient than G2, which means
\textgreater 45 \% in $J$-band (1.310 $\mu$m).
\item[(B)] Field dependent efficiency requirement: The
efficiency of the new grating has to be better than G2 efficiency
(\textless 30 \%) at all location across the
grating. 
\end{itemize}
In the next section, we report the results of the new grating efficiency measurements.

\subsection{New (G3) and Old (G2) OSIRIS Grating Efficiencies}
Before G3 was shipped to the Dunlap Institute
for Astronomy \& Astrophysics
(Dunlap) in July 2012, Bach Research assessed the quality
of the wavefront of G3 surface
with a 4 inch aperture Zygo interferometer. Bach Research took wavefront measurements
along the center of the grating and moved the aperture from start to the
end of the ruling. They performed these measurements across several
optical orders to yield an indication of wavefront error over the entire
surface. The wavefront quality across
the surface is roughly $\pm$ 0.5 wave at $\lambda$ = 632.8 nm.  

\begin{deluxetable}{cccc}
\tablecolumns{4}
\tablewidth{0pc}
\tablecaption{G3 Peak Efficiency at 1.310 $\mu m$}
\setlength{\tabcolsep}{0.07in}
\tablehead{
\colhead{Location} & \colhead{Left} &
\colhead{Center} & \colhead{Right}
}
\startdata
    & 82.5 $\pm$ 5.1 \%    & 76.8 $\pm$ 4.7 \% & 80.4 $\pm$ 5.0 \% \\
Top & 78.2\tablenotemark{a} & 77.9 $\pm$ 4.2 \% & 79.2 $\pm$ 4.5 \% \\ 
    & 1.05                 & 0.99              & 1.01 \\
\tableline
       & 76.9 $\pm$ 4.8 \% & 76.7 $\pm$ 4.7 \% & 79.3 $\pm$ 4.9 \% \\
Center & 76.4 $\pm$ 4.3 \% & 76.3 $\pm$ 4.3 \% & 79.4 $\pm$ 4.4 \% \\
       & 1.00              & 1.01              & 1.00 \\
\tableline
       & 77.6 $\pm$ 4.8 \% & 75.2 $\pm$ 4.6 \% & 77.9 $\pm$ 4.8 \% \\
Bottom & 77.6 $\pm$ 4.3 \% & 76.1 $\pm$ 4.1 \% & 79.3 $\pm$ 4.3 \% \\
       & 1.00              & 0.99              & 0.98 \\
\enddata \label{eff_table_new}
\tablenotetext{NOTE}{At each location, there are three values. The
  first one is the efficiency with respect to the pure reflection, the
  second one is the efficiency with respect to the order sum, and the
  last one is the ratio of the order sum to the pure reflection. All
  values are shown with the associated measurement errors.}
\tablenotetext{a}{After we reduced all data, one of the original data
  file (m = 6 at TL) was corrupted before we calculated the random
  observation uncertainty. Thus, the efficiency value is reported here,
  but not the measurement uncertainty.}
\end{deluxetable}

\begin{figure*}[!h]
\begin{center}
\includegraphics[width=1\textwidth]{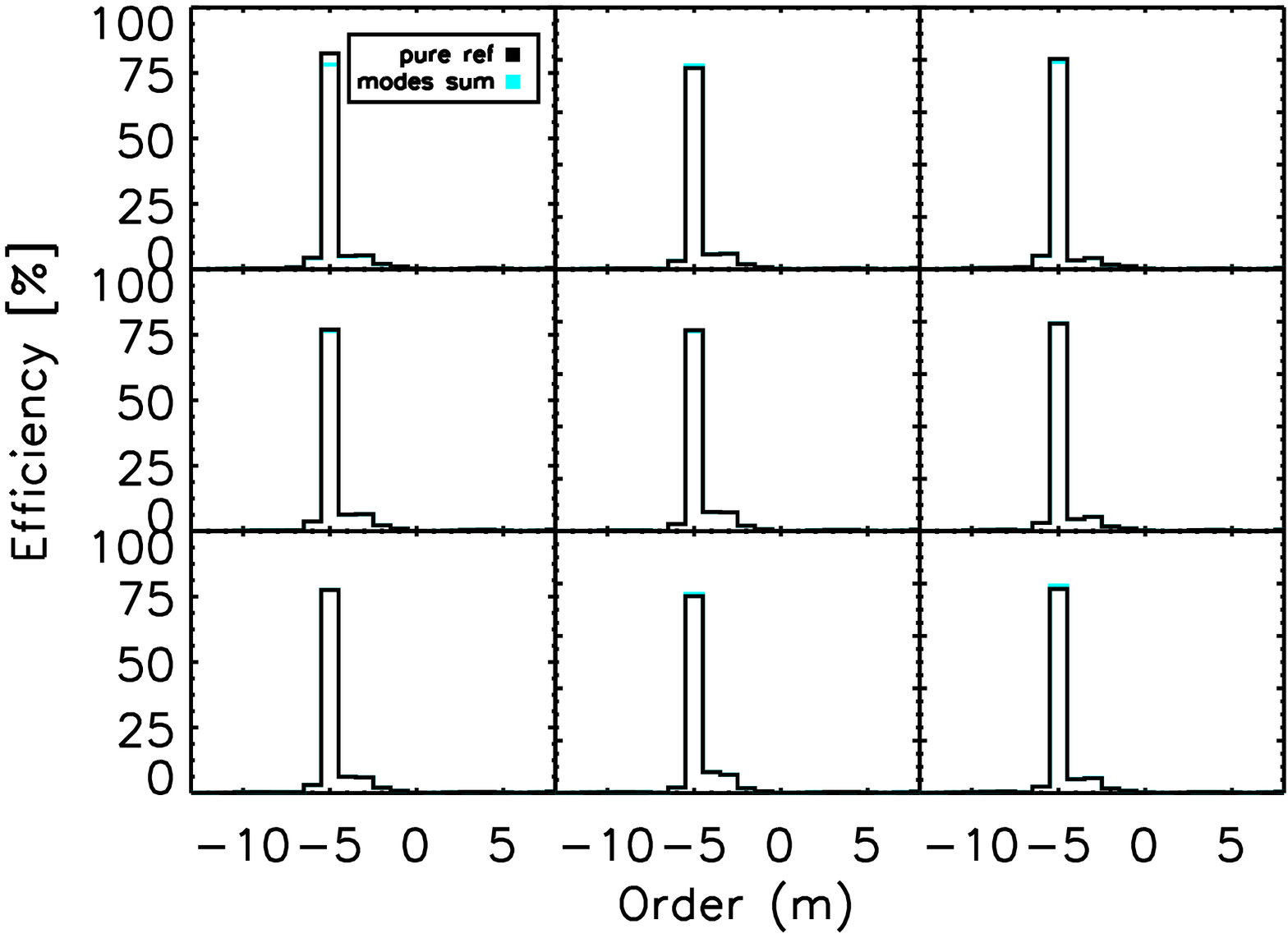}
\caption{Efficiency of G3 by Bach Research with
  respect to the pure reflection (black) and with respect to the
  order sum (cyan) measured at -13
  $\leq$ $m$ $\leq$ 8.}
\label{third_eff}
\end{center}
\end{figure*}

The peak efficiency ($m = -5$) of G3 with respect to the pure
reflection at $\lambda$ = 1.310 $\mu$m at nine
spatial locations across the grating surface are summarized in Table
\ref{eff_table_new},
and Figures \ref{third_eff} shows detailed efficiency for -13
$\leq$ $m$ $\leq$ 8.
At all nine locations, the peak efficiencies are more than 75 \%, and
the average efficiency, weighted by error, is 78.0 $\pm$ 1.6 \% with
respect to the pure reflection, and the non-weighted average of 77.8 \% with respect to the
order sum (see footnotes in Table \ref{eff_table_new}). This
grating meets both global and field dependent efficiency requirements stated in
\textsection \ref{procedure}, which led
our team to install G3 in OSIRIS in December 2012.

\label{section_old}
After G3 was installed and its on-sky performance was confirmed through
engineering observations, G2 was shipped to Dunlap,
and its efficiency was measured using the same setup used for G3.
\begin{deluxetable}{cccc}
\tablecolumns{4}
\tablewidth{0pc}
\tablecaption{G2 Peak Efficiency at 1.310 $\mu m$}
\setlength{\tabcolsep}{0.07in}
\tablehead{
\colhead{Location} & \colhead{Left} &
\colhead{Center} & \colhead{Right}
}
\startdata
 & 64.6 $\pm$ 4.0 \% & 30.6 $\pm$ 1.9 \% & 31.1 $\pm$ 1.9 \% \\
Top & 59.5 $\pm$ 3.4 \% & 28.7 $\pm$ 1.5 \% & 29.4 $\pm$ 1.5 \% \\
    & 1.09              & 1.06              & 1.06 \\
\tableline
       & 54.8 $\pm$ 3.4 \% & 50.6 $\pm$ 3.1 \% & 34.9 $\pm$ 2.2 \% \\
Center & 45.3 $\pm$ 2.8 \% & 41.5 $\pm$ 2.6 \% & 32.9 $\pm$ 1.7 \% \\
       & 1.21              & 1.22              & 1.06 \\
\tableline
       & 49.9 $\pm$ 3.1 \% & 47.0 $\pm$ 2.9 \% & 37.9 $\pm$ 2.3 \% \\
Bottom & 46.3 $\pm$ 2.5 \% & 43.4 $\pm$ 2.3 \% & 35.9 $\pm$ 1.8 \% \\
       & 1.08              & 1.08              & 1.06 \\
\enddata \label{eff_table_old}
\tablenotetext{NOTE}{At each location, there are three values. The
  first one is the efficiency with respect to the pure reflection, the
  second one is the efficiency with respect to the order sum, and the
  last one is the ratio of the order sum to the pure reflection. All
  values are shown with the associated measurement errors.}
\end{deluxetable}

\begin{figure*}[!h]
\begin{center}
\includegraphics[width=1\textwidth]{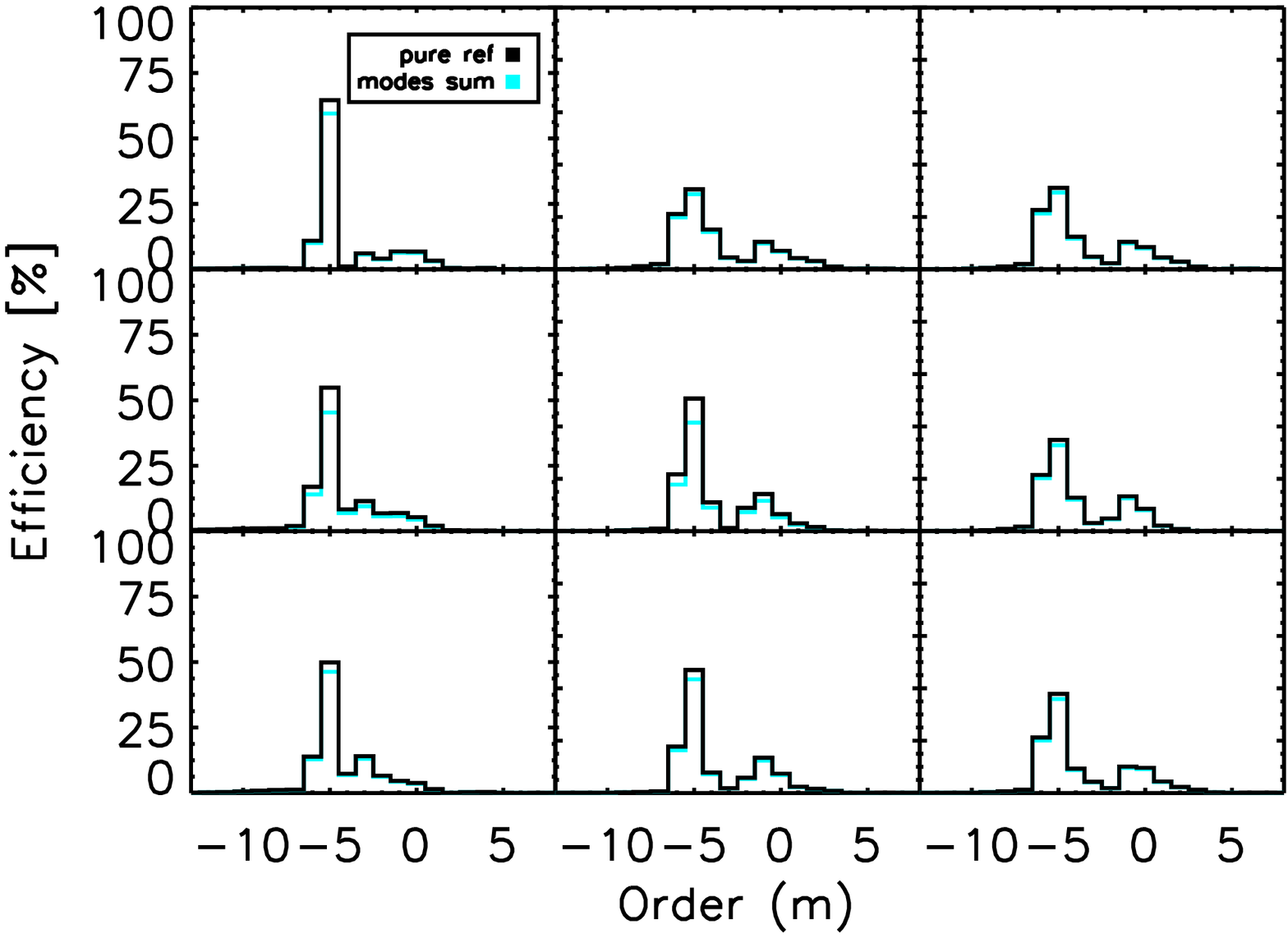}
\caption{Efficiency of G2 by Diffraction
  Products with respect to the pure reflection (black) and with respect
  to the order sum (cyan) measured at -13 $\leq$ $m$ $\leq$ 8.}
\label{second_eff}
\end{center}
\end{figure*}
Table
\ref{eff_table_old} summarizes the peak efficiency at nine locations,
and Figure \ref{second_eff} is the result of full efficiency
measurements of G2. On average, G2 has a weighted efficiency of 39.5
$\pm$ 0.8 \% with respect to the pure
reflection and 35.8 $\pm$ 0.7 \% with respect to the order sum.

On average, G3 has a factor of about two greater
efficiency at 1310 $\mu$m. We also find
that G3 has a close-to-uniform efficiency across the
surface compared to G2. The field-dependent standard
deviation of G3 peak efficiencies is
2.23 \% whereas the field-dependent standard deviation of G2
is 11.68 \%.

\subsection{Measurement Uncertainties}
To estimate the uncertainties in the efficiency measurement, we look at
the configuration uncertainty of 1\% and flat fielding
uncertainty of 2.4 \%  (\textsection \ref{procedure}),
random camera noise of $\sim$ 10 DN  (Appendix
\ref{sec_camera}), flux fluctuation due to PCB
temperature of 0.3 \% (Appendix \ref{sec_camera}) and laser
stability of 3.5\% (Appendix \ref{sec_laser}). 
We also estimate the
random observational error by calculating the pixel-wise standard
deviation of the mean using 100 science frames and 100 dark frames per
order. The random observational error is very small ($\textless$ 0.02
\%) for all cases.

To confirm the repeatability of our measurements and the estimate of the
error, we measured G2 efficiency at two
locations, top-left and top-center, (see Figure \ref{eff_loc} for the
location) twice, the second measurement after about two weeks later than
the first measurement.
The peak efficiency ratio of the first time to the second time is 0.982
(pure reflection) and 0.995 (order sum) for the top-left location, and 1.005
(pure reflection) and 1.011 (order sum) for the top-center location. They are both
within the measurement uncertainties.

\subsection{Polarization Effect on Grating}
\label{sec_polar}
The polarization state of the incident light can affect the efficiency
of a diffraction grating in many cases. To understand the polarization dependence on the
OSIRIS grating efficiency, we
modelled the TE (polarized parallel to the groove) and TM (polarized
perpendicular to the groove) efficiencies of the OSIRIS grating using
RCWA.
The RCWA model predicts that the peak TE to TM efficiency ratio for a 1.310
$\mu$m monochromatic light source at $m = -5$ is 1.041. 

\begin{figure}[!h]
\plotone{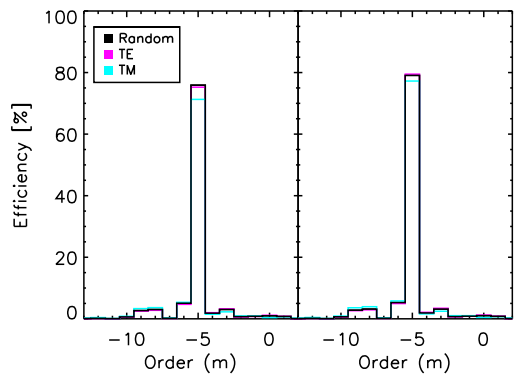}
\caption{Efficiency of the test ruling by Bach Research with respect to the
  pure reflection (left) and with respect to the order sum
  (right) measured (-13 $\leq$ $m$ $\leq$ 2) using
  TE/TM modes (magenta/cyan) and no
  polarizer (black).}
\label{test_eff_pureref_tot}
\end{figure}

We conducted the polarized efficiency measurement experiments on the
test ruling three times using: (1) the IR
laser diode (no polarizer); (2) the IR laser diode in TE mode defined by
the polarizer; and (3) the IR laser diode in TM mode defined by the
polarizer. During the measurement, all the components, especially the
fiber was fixed to keep the same polarization state throughout.
Figure
\ref{test_eff_pureref_tot} shows these results where the total
efficiency is defined either by the pure reflection (left) or the order
sum (right). 
The peak TE to TM efficiency ratio at $m = -5$ is 1.056 $\pm$ 0.092 and
1.029 $\pm$ 0.074 for the
pure reflection and the order sum, respectively. These
measurements are in agreement with the theoretical
predictions.

We confirmed that
the polarization state of the laser does not change
during the full efficiency measurement at one spatial location
(\textsection \ref{procedure} and Appendix \ref{sec_laser}), and thus
individual orders and the pure reflection are measured with the same
polarization. As the efficiency is calculated with respect to the pure
reflection or the order sum, we do not take into account the effect of
the polarization in the measurement uncertainty calculation. Between
different spatial locations, maximum of 4\% difference in efficiency due
to TE and TM states can theoretically occur.
This polarization effect can also affect the real scientific observation at
OSIRIS; however, it is probably insignificant since other noise
components, such as sky lines, would
be the dominant source of uncertainty. 

\section{New Grating Installation and Commissioning}
\label{installation}
OSIRIS was slowly warmed up to ambient temperature over a one week period, and 
G3 installation was performed in December 2012 by our team. G2 was
removed from OSIRIS, and its alignment was measured and
marked with respect to the mounting plate in the lab at the Keck
Observatory summit facility. G2 was
detached from the mounting plate, and G3 was aligned to
the marks and installed on the mounting plate. 

\begin{figure}[!h]
\plotone{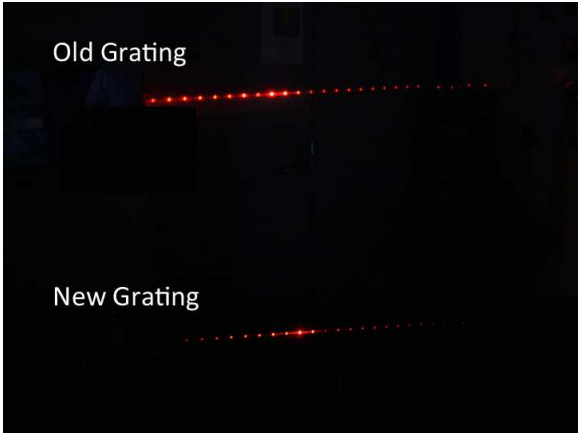}
\caption{Diffraction of a HeNe laser at 632.8 nm using G2 (top) and G3
  (bottom). Test performed in December 2012. The light is more
  concentrated in one order for G3 whereas high fraction of light is
  diffracted in multiple orders for G2.}
\label{second_third_grating}
\end{figure}

Figure
\ref{second_third_grating} shows a HeNe laser at 632.8 nm being diffracted by G2
(top) and G3 (bottom) in the Keck summit
lab prior to the installation of G3 in OSIRIS. The light is visibly
more concentrated to one order for G3 while a higher fraction of light
is diffracted to multiple orders by G2.

On January 20 and 27, 2013, OSIRIS with G3 was commissioned on-sky using
the Keck-I AO system. Both
nights started with clear night, but unfortunately, within one
or two hours, the weather conditions changed to thin/high cirrus with
high wind. Some standard stars and blank sky were observed with the deformable
mirror off. 
The resulting measurements are certainly affected by the varying weather
conditions.
The measurements of OSIRIS sensitivity with G3 in each spectrograph
filter from these nights, as well as prior measurements with G2 are
reported in \textsection \ref{onsky}.

\section{On Sky Performance and Throughput}
\label{onsky}
In this section, we compare on-sky performance between G2 and G3. 
Our comparison is complicated by the fact that in early 2012 OSIRIS was
moved from Keck-II to Keck-I to be the first
dedicated science instrument for a new Laser Guide Star (LGS) AO
capability on Keck-I that was installed in 2010. The LGS system on Keck-I uses a
significantly improved
laser system compared to the existing Keck-II laser system
\citep{chin:10}. 

The final zero point magnitude at all broadband filters for OSIRIS are
calculated using standard star observations.
The data used to calculate G2 zero points were all taken on
Keck-II, and G3 data were all taken on Keck-I. Although there are
differences in throughput between the two telescopes and AO systems as well as
differing weather condition between our limited observations,
there is a general zero point improvement for G3 in comparison to G2. 

We use as many standard star observations as we had
access to, but the number of observations is fairly small in each
particular band. For OSIRIS with G2 on Keck-II, HD105601, HD106965, HD201941,
and HD18881 taken between April 2007 and January 2012 are
used, and for OSIRIS with G3 on Keck-I, HD44612 and HD18881 observed on
January 20 and 27, 2013 are used. 
The zero points are calculated by applying a large rectangular
aperture on raw (non-reduced) image, over the entire spectrum. In all cases,
approximately equal rectangles are used for the wavelength ranges.
The resulting zero point magnitude and the factors of
improvement are shown in Table \ref{zeropoint}.
\begin{deluxetable}{cccccc}
\tablecolumns{6}
\tablewidth{0pc}
\tablecaption{Zero Point Vega Magnitude and Factor of Improvement from
  G2 to G3}
\setlength{\tabcolsep}{0.07in}
\tablehead{
\colhead{Filter} & \colhead{$Zbb$} & \colhead{$Jbb$} & \colhead{$Hbb$} &
\colhead{$Kcb$} & \colhead{$Kbb$}\\
\colhead{$\lambda_{cen}$ [$\mu$m]} & \colhead{1.090} & \colhead{1.325} &
\colhead{1.637} & \colhead{2.174} & \colhead{2.174} 
}
\startdata
G2 at Keck-II & 23.70 & 23.80 & 24.45 & 23.19 & 23.67 \\
G3 at Keck-I & 24.19 & 24.22 & 24.84 & 24.16 & 24.54 \\
Improvement & 1.56 & 1.48 & 1.43 & 2.45 & 2.23 \\
\enddata \label{zeropoint}
\end{deluxetable}

\section{Conclusion}
OSIRIS at W. M. Keck Observatory is a particularly unique IFS instrument
among other IFSs with AO
capability today in use of a single fixed exceptionally coarse ruling (27.93
grooves per mm) diffraction grating, which uses $m = -3$, $-4$, $-5$, and $-6$
to cover $K$, $H$, $J$, and $Z$ bands. While OSIRIS has delivered a
number of important scientific results, its sensitivity was
limited by the performance of its spectrograph grating. Our team has
worked with a new grating vendor, Bach Research Corporation, to produce
a better quality grating for OSIRIS.

Bach Research manufactured a test ruling and the new grating (G3) in 2012, and we have
carefully measured the direct efficiencies of both at 1.310
$\mu$m in the lab. The weighted field-averaged
peak efficiency of G3 is 78.0 $\pm$ 1.6 \% (pure reflection) and
77.8 \% (order sum) (see footnotes in Table \ref{eff_table_new}) with a field
standard deviation 2.23 \% (pure reflection) and 1.32 \% (order sum). After
the G3 efficiency was
confirmed to be high and close to uniform over the surface, G3 was
installed to OSIRIS in
December 2012. G2 was shipped to Dunlap after G3
performance was tested and confirmed on-sky in January 2013. G2
efficiency was as well measured using the same lab setup used for the G3
measurement. For G2, the weighted field-averaged peak efficiency is 39.5 $\pm$ 0.8 \% (pure
reflection) and 35.8 $\pm$ 0.7
\% (order sum) with a field standard
deviation is 11.68 \% (pure reflection) and 9.82 \% (order sum). 

The new OSIRIS grating gives a factor of about two times
increase in average efficiency at 1.310 $\mu$m with less
field-dependent efficiency change
across the surface. The final sensitivity improvement was difficult
to assess because OSIRIS was moved from Keck-II to Keck-I in early 2012;
however, we were able to determine the zero point magnitudes and factors
of improvement for each broadband
filter. On average, on-sky throughput is 1.83 times better than when it
was at Keck-II with G2. This enables us to observe fainter
objects and to use observing time more efficiently.

A single fixed diffraction grating with a coarse ruling can reach high
efficiency and perform well on OSIRIS, but it is very difficult to
engineer, and only a few companies exist worldwide who can manufacture
such a grating today. For the
next generations of IFS instrumentation, the more usual approach of
using finer ruling grating with $m = 1$ order may be better suited. For
example, IRIS on TMT, an IFS with some characteristics and design
elements similar to OSIRIS, will instead have several gratings with finer groove
densities of $\sim$ 150 to 900 per mm \citep{moore:10, larkin:10}. 

\section*{Acknowledgements}
We particularly want to thank the Bach Research Corporation for all of
their efforts and
support during the selection and fabrication of the OSIRIS grating.  We
enjoyed working with the Bach Research Corp. team during this entire
process. We also thank the Keck Observatory staff who helped with the
planning and installation work for the new grating in OSIRIS. The new
OSIRIS grating was graciously funded
by the Dunlap Institute for Astronomy \& Astrophysics at University of
Toronto. Funding for this project was also provided by the NSERC
Discovery grant (RGPIN 419376) and the Canada Foundation for Innovation
grant (31773). Our on-sky data presented herein were obtained at the
W.M. Keck Observatory, which is operated as a scientific partnership
among the California Institute of Technology, the University of
California, and the National Aeronautics and Space Administration. The
Observatory was made possible by the generous financial support of the
W.M. Keck Foundation. The authors wish to recognize and acknowledge the
very significant cultural role and reverence that the summit of Mauna
Kea has always had within the indigenous Hawaiian community.  We are
most fortunate to have the opportunity to conduct observations from this
mountain. 

\appendix
\twocolumn
\section{Camera Linearity}
\label{sec_camera}
\begin{figure}[!h]
\plotone{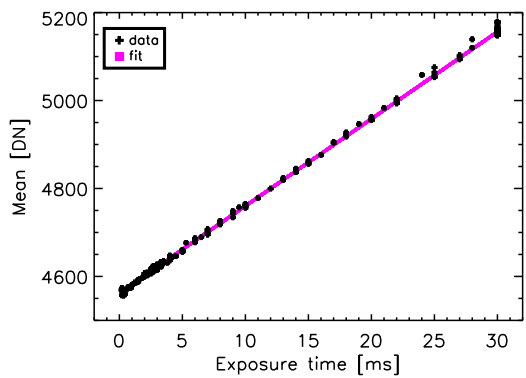}
\caption{Average dark counts vs. exposure time at the time of efficiency
  measurements. The data were fit by a straight line shown over plotted
  in magenta.}
\label{linearity_test0}
\end{figure}

We evaluated the linearity of the InGaAs camera, Raptor Photonics OWL SW
1.7 CL-320, by measuring the average dark
counts on the detector as a function of exposure time. The 100 dark
frames per grating order taken at the time of efficiency measurements
were median combined to make the master dark, and the average of the
master dark was plotted as a function of exposure time on
Figure \ref{linearity_test0}.
All camera settings were kept fixed
for all times except for the exposure time.
\begin{figure}[!h]
\plotone{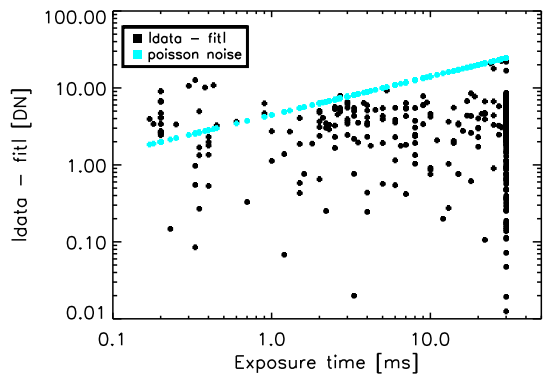}
\caption{Absolute difference between the average dark count and the
  linear fit plotted as a function of exposure time, with the Poisson
  noise overplotted in cyan.}
\label{linearity_test1}
\end{figure}
Figure \ref{linearity_test1} shows the absolute difference between data
points and the straight line fit, with the Poisson noise over plotted in cyan.
The result shows that regardless of the exposure time, there are about
10 DN fluctuations until the Poisson noise takes over at
around $t_{exp}$ = 10 ms. Hence, we include 10 DN in the noise calculation.

We also looked at the effect of the camera temperature over an
hour, which is the amount of time taken for an efficiency measurement of a single
spatial position on the grating surface over a large range
of orders. In this test, we took images of the laser spot at $t_{exp}$ = 0.2 ms
every 60 seconds for one hour and repeated the process twice. During the
test, the camera and the laser were kept on all the time. The laser was imaged
to include the effect of the laser heating up the image sensor in case that happened.
\begin{figure}[!h]
\plotone{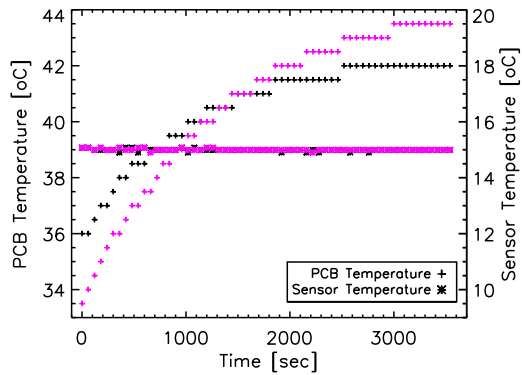}
\caption{Sensor and PCB temperature changes over one hour. With the TEC
  temperature set to 15 C$^\circ$, the sensor temperature stayed almost
  constant while the PCB temperature increased 5 to 10$^\circ$C. The
  magenta points are the measurements from the first test, and the black
  points are the measurements from the second test.}
\label{temp_time}
\end{figure} 
Figure \ref{temp_time}
shows the changes of two different camera temperatures over one hour
period. Our input temperature 
(TEC temperature) was always 15$^\circ$C, and the sensor temperature
was almost always constant around 15$^\circ$C, but the printed circuit board
(PCB) temperature increased about 10$^\circ$C during the first test and
about 6$^\circ$C during the second test over a one hour period.

\begin{figure}[!h]
\plotone{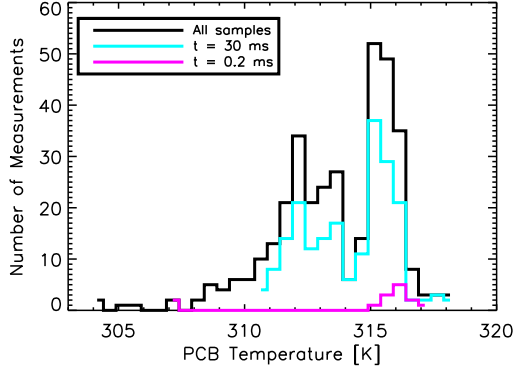}
\caption{The distribution of PCB temperature during G2 and G3
  efficiency measurements combined. Black line is the
  distribution among all samples whereas
  cyan and magenta are among t = 30 ms and t = 0.2 ms respectively.}
\label{sample_hist_temp}
\end{figure}

\begin{figure}[!h]
\plotone{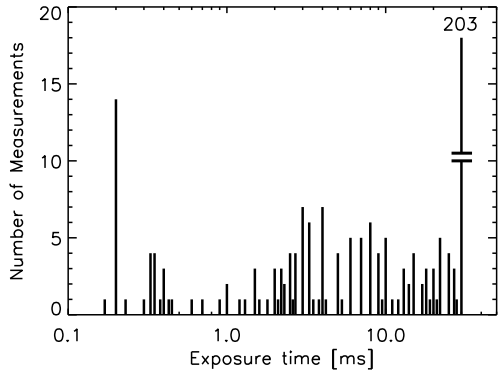}
\caption{The distribution of measurements in an exposure time space
  during G2 and G3 efficiency measurements combined.}
\label{sample_hist_expt}
\end{figure}

The distribution of PCB temperatures and the distribution of exposure
times used during G2 and G3 efficiency measurements
combined is shown in Figure \ref{sample_hist_temp}, and Figure
\ref{sample_hist_expt}. 
Samples taken at $t_{exp}$ = 30 ms are
dominant because G3 and G2 efficiency measurements were all taken with a
frame rate of 25 Hz, and taking into
account the trigger delay and data transfer, we set the maximum exposure
time to be $t_{max} = 30$ ms. The exposure times for individual
measurements were chosen to maximize signal level while maintaining the
exposure below the saturation, but since we set a maximum exposure time
threshold, many fainter spot images were taken
with the maximum exposure time. 

\begin{figure}[!h]
\plotone{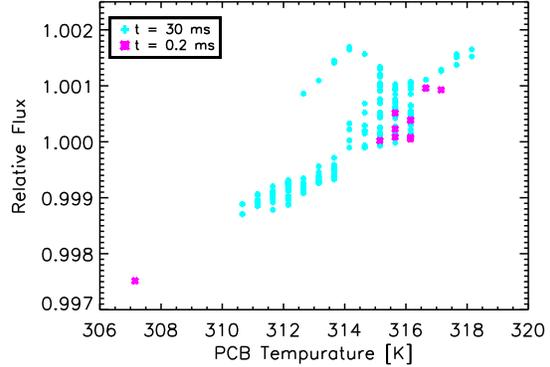}
\caption{The relative flux change of t = 30 ms (cyan) and t = 0.2 ms
  (magenta) average dark counts due to the PCB temperature change.}
\label{sample_hist}
\end{figure}

Figure \ref{sample_hist} shows average dark counts, normalized by the
average of all the average dark counts, as a function of the PCB
temperature change.
The flux increases about 0.6 \% as the PCB
temperature increases about 10$^\circ$C. We do not have
the information on how long the camera was turned on during the experiments, but we know that to
measure the full efficiency at one location, it takes about one hour, and in
one hour the PCB temperature changes about 10$^\circ$C (Figure
\ref{temp_time}). Combined this information and Figure
\ref{sample_hist}, we take 0.3 \% (a half of the full increase in flux)
as noise due to camera PCB temperature changes during the efficiency measurements.

\section{Infrared Laser Diode Stability Test}
\label{sec_laser}
\begin{figure}[!h]
\plotone{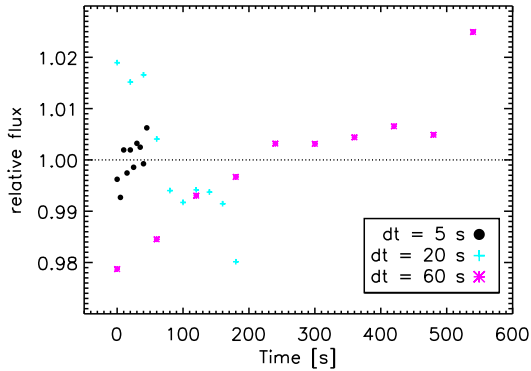}
\caption{Relative flux of the 1.310 $\mu$m laser diode verses time for three
  time intervals: 5 (black circle), 20 (cyan cross), and 60 (magenta star) seconds.}
\label{total_by_mean}
\end{figure}

We used a 1.310 $\mu$m laser diode coupled to a SMF-28 fiber as a
light source for infrared measurement. To
ensure consistency in our measurements, we measured the stability of the
laser by monitoring its intensity using the infrared camera tested in
Appendix \ref{sec_camera}. First, we
fixed the exposure time to $t_{exp}$ = 0.2 ms and took a series of images separated by three
time intervals: 5, 20, and 60 seconds (Figure \ref{total_by_mean}). 
The
camera and the laser were kept being on until all measurements in a
particular sequence were completed. These measurements
allowed us to search for any time dependent instabilities in the
combined system of laser and camera. We found the fluctuation in the
system seems independent of the time interval of data taken but
dependent on the duration of the laser is on.

The polarization of stimulated emission is parallel to the diode
junction plane, and thus laser diodes are usually linearly
polarized. When a laser is used in a
polarization dependent setup, intensity fluctuations can occur due to
changing polarization states. Our laser diode is coupled to a SMF-28-J9
step-index fiber with a numerical aperture (NA) of 0.14 and an 8.2 $\mu$m
diameter core. For this fiber, the dimensionless normalized frequency or
normalized thickness of the guide ($V$) is,
\begin{equation}
V = k d (NA),
\end{equation}
where $k$ is the wave number, and $d$ is the fiber core radius
\citep[e.g.][]{iizuka:02} of 2.75
at 1310 $\mu$m. The first critical frequency (cutoff $V$) for
single mode operation is
2.405, and therefore our fiber supports four polarization
modes. To understand the polarization characteristics of our laser, we
tested the laser stability with and without
a calcite polarizer in the optical path. We took a series of images of
laser beam for an hour at 60 s intervals. We performed these
measurements twice: once with a calcite polarizer in front of the laser
and once without the polarizer. The polarizer was
oriented so that the output beam had the maximum intensity (direction of
the polarization of the laser is parallel to the direction of
polarizer). After one hour of measurement, we verified that the peak
intensity from the polarizer was still at the same angle. This means,
over one hour, the polarization state of the laser did not change, and
therefore we assume the laser polarization state does not change during the
efficiency measurements.

\begin{figure}[!h]
\plotone{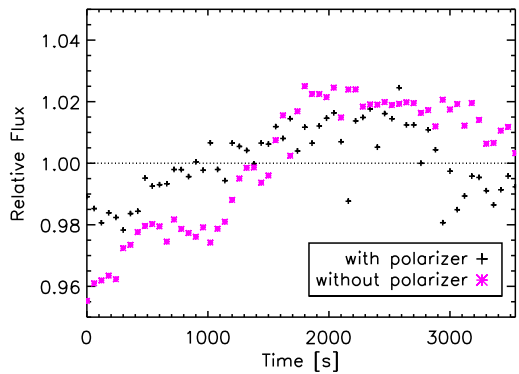}
\caption{Relative flux of the 1.310 $\mu$m laser diode versus time for a 60
  second interval with a polarizer (black) and without a polarizer
  (magenta).}
\label{pol_no_pol}
\end{figure}

Figure \ref{pol_no_pol} shows that the laser flux fluctuates 3.5
\% (one-half of (highest - lowest) flux) over one hour. 
The initial discrepancy up to 1000 s between the measurements with and
without the polarizer is probably due
to room/detector temperature differences because the two measurements were
taken on different days. Since the
polarization state of the laser did not change
over one hour, this variation is probably from the laser itself as shown
in Figure \ref{total_by_mean}. We take into account this 3.5 \% laser flux
variation in the noise calculation.

\section{Total Flux in a Spot}
\label{justification}
\begin{figure}[!h]
\plotone{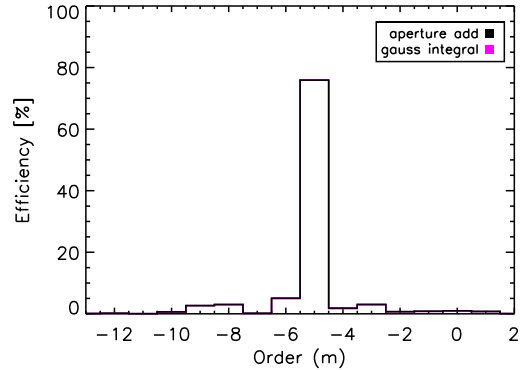}
\caption{Efficiency of the test ruling with respect to the pure
  reflection. For the two cases, a circular aperture applied to the reduced image
  (black) and Gaussian fit integral (magenta), the efficiencies are
  almost on top of each other.}
\label{apr_gau}
\end{figure}

The grating efficiency is measured by summing up the flux in an
individual spot at a particular order and dividing by the total flux. To
confirm that we  collect all flux in an individual spot, we tested two
methods. One is by Gaussian fitting. A two dimensional Gaussian function
is fitted to the final reduced image, and aperture photometry is
applied to both the reduced image and the 2D
Gaussian function.
\begin{figure}[!h]
\plotone{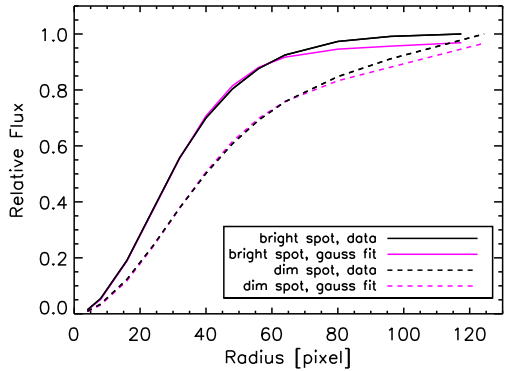}
\caption{Growth curves of the reduced image (black) and 2D Gaussian
  fit (magenta) for the test ruling. The solid lines are for the brightest
  spot, and the dashed lines are for a dim spot whose optical path is the
  longest whose size is therefore the biggest on the detector.}
\label{gaussfit_argue}
\end{figure}
Figure \ref{gaussfit_argue} shows the growth curves
of the reduced image (black) and the Gaussian (magenta) for the test
ruling. They are both normalized to the reduced image total flux.
The solid line is the growth curve of the brightest spot, and the dashed
line is of a dimmer spot. The optical path of the dimmer spot is the
longest and thus the spot size is biggest on the detector due to the
divergence angle of the laser beam.
The maximum radius on the plot 
is the radius of a biggest circle that can be fitted in the frame centered
at the Gaussian fit center. For both bright and dim spots, two lines are
quit similar.
Figure \ref{apr_gau} shows the efficiency of the test ruling with
respect to the pure reflection, using the
aperture photometry of the reduced image (black) and the integrated sum of the
Gaussian to infinity (magenta).
Two plots are almost exactly the same.

The growth curve and efficiency comparisons illustrate that our
experiment and optical setup is optimized with respect to the laser spot
size at the detector and the detector plate scale. Since two cases give
similar answers, we deploy the simple aperture photometry method to the
reduced image to calculate the total counts in a spot. 

\section{Pipeline Modification}
\label{pipeline}
One unique aspect of OSIRIS is that over 3000 spectra are all partially
overlapped on the detector at staggered wavelength, and hence special
reduction and calibration steps are required. The OSIRIS Data Reduction
Pipeline (DRP) reduces the science data to
the level where a user can begin their custom scientific analysis. 
After OSIRIS was moved
to Keck-I, and G3 was installed, the DRP had to be
modified to account for the new AO system and the new grating.

On Keck-I, the AO system optical path to OSIRIS has one less mirror than
the Keck-II system. This produces a flipping of image in the
y-direction. This axis flip is fixed in the DPR ``Assemble Data Cubes''
module. The Keck-I AO system uses a different IR/visible splitting
dichroic from that used on Keck-II, and white light measurements using
the AO fiber calibration source showed that the Keck-I AO system dichroic produces
essentially no instrumental dispersion. The ``Correct
Dispersion'' module in the DRP that corrects
for atmospheric dispersion and instrumental dispersion was appropriately modified
for the Keck-I AO path.

\begin{figure*}[!h]
\begin{center}
\includegraphics[width=1\textwidth]{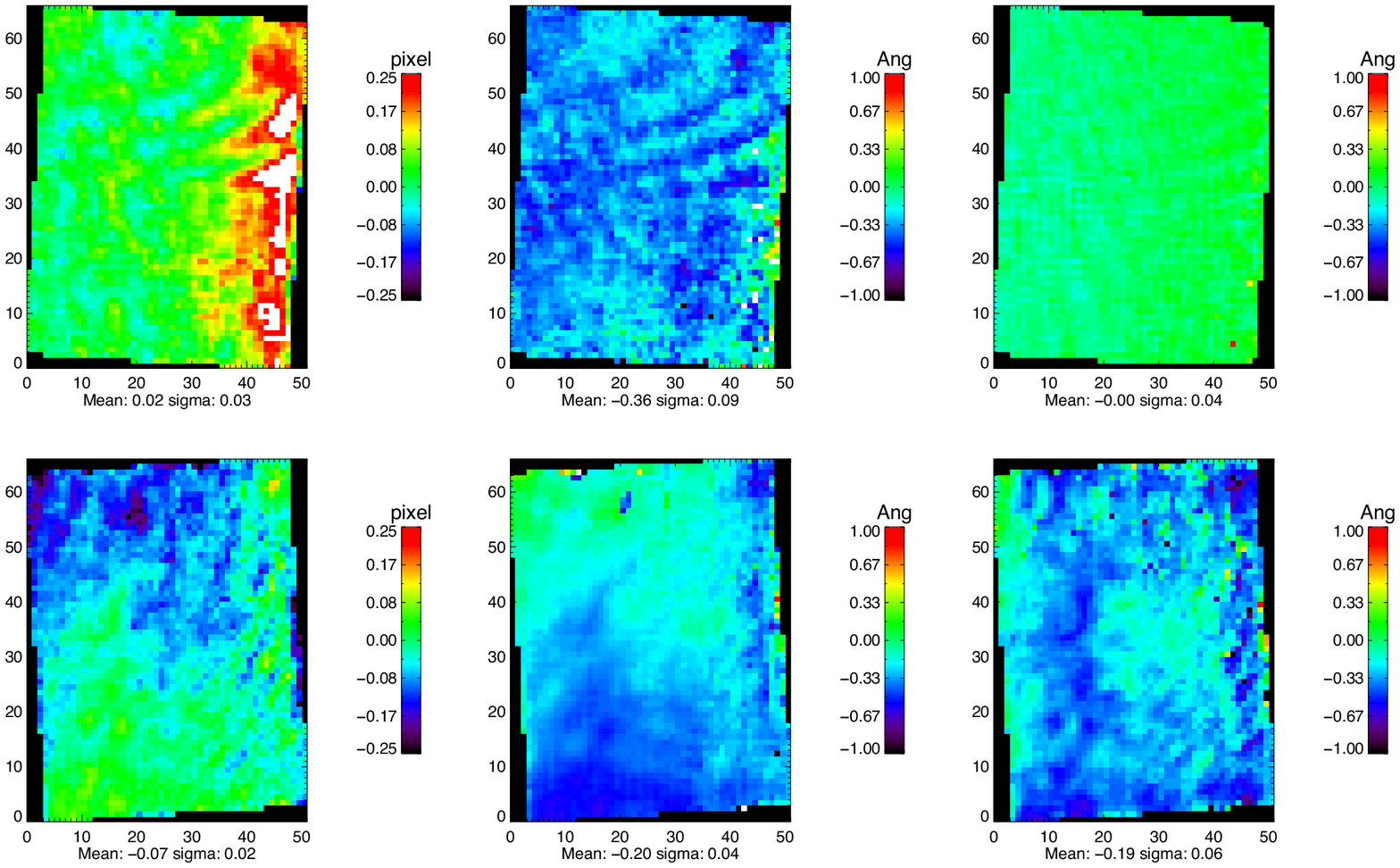}
\caption{Uncorrected relative pixel shift (left), uncorrected
  absolute angstrom offsets (middle), and corrected absolute
  angstrom shift (right) for Kn3 35 mas in 2006 with G2 (top) and in 2013
  with G3 (bottom).}
\label{wav_sol}
\end{center}
\end{figure*}

We determined a new wavelength solution using arc lamp and OH lines in
each broadband filter, and we found about four spectral channels of shift from the
previous (G2) version. The field dependent
wavelength solution per spaxel was calculated
using the cross correlation of OH lines using the Kn3 filter with the 35
mas spaxel scale. Figure
\ref{wav_sol} shows relative/absolute pixel and angstrom wavelength
offsets between 2006 with G2 (top) and 2013 with G3 (bottom).
It is a
residual offsets after using the ``Assemble Data Cubes'' and a new
wavelength solution.
The comparison of two confirms that the quality of the ruling on the new
grating is more uniform across the surface. The field
dependent and global wavelength solution is now implemented in the new
pipeline. 

A problem had been observed prior to commissioning of G3 that affected
the lower right quadrant of the reduced
data cube with varying shifts in measured intensity and shifts in the
detector channel offsets. During the modification of the DRP, we found
that there appeared to be a bad column on the detector, and this was
biasing the wavelength solution. This problem started on September
17, 2011 when one of the spectrograph Hawaii-II detector Leach-ARC
detector read-out video boards was swapped. In the end, it
is just one pixel shift that skews the timing of the entire
channel. This channel offset is now fixed in the new pipeline within the
``Subtract Frame'' module.

All these modifications are implemented in the new version of the DRP,
and the DRP full package is now available to download at the OSIRIS instrument
webpage\footnote{http://www2.keck.hawaii.edu/inst/osiris/}.

\bibliographystyle{apj}
\bibliography{reference}

\end{document}